# A common interface for multi-rule-engine distributed systems

Pierre de Leusse, Bartosz Kwolek and Krzysztof Zieliński

Distributed System Research Group, AGH University of Science and Technology
Krakow, Poland
{pdl, bkwolek, kz}@agh.edu.pl

**Abstract.** The rule technological landscape is becoming ever more complex, with an extended number of specifications and products. It is therefore becoming increasingly difficult to integrate rule-driven components and manage interoperability in multi-rule engine environments. The described work presents the possibility to provide a common interface for rule-driven components in a distributed system. The authors' approach leverages on a set of discovery protocol, rule interchange and user interface to alleviate the environment's complexity.

## 1 Introduction

In recent years, rule-based systems have become increasingly popular [1]. This evolution has been mostly attributed to three factors, 1) better separation of concerns between the knowledge and its implementation logic in contrast to a hard-coded approach; 2) rule repositories that increase the visibility and readability of the knowledge and 3) graphical user interfaces that render rules more usable while bridging the gap between users (e.g. domain experts) and IT specialists.

Influenced by this increased interest, the technological landscape of rules is becoming more and more complex. This is partly due to the number of technologies being developed and the frequency in which they appear. In particular, the amount of platforms implemented (e.g. Oracle Business Rules [2], Drools [3], Jess [4]) as well as the various specifications related to rule expression and enactment (e.g. RuleML [5], RIF [6]) have rendered this domain more opaque.

This abundance of technologies and products can be beneficial as different approaches attempt to address a variety of problems (e.g. production, reaction). However, it greatly impacts the usability of distributed systems that leverage on rule engines in order to automate managed components behavior. The behavioral and functional complexity reduced by rule engines at the component level (e.g. OSGI Bundle [7]) translates into management and interoperability issues in the distributed application plane. Indeed, rule visibility and interchange as well as reasoning interoperability have become a new challenge in the implementation of business logic.

The main contribution of this paper and the demonstration system it presents, is a common user interface for multi-rule-engine distributed systems. This demonstration addresses challenges in the domains of rule-driven component discovery, rule interchange and multi-rule-engine usability.

The authors have implemented a web user interface along with supporting tools that allow discovering rule engines in a distributed environment, retrieving from as well as registering rules to them and interchanging rules between them. This approach aims at providing an integration of different rule engine technologies as seamlessly as possible while allowing an improved user experience with a rich interface.

This paper is organized as follows: Section 2 presents the requirements for a common interface in multi-rule-engine distributed systems. Section 3 presents the anatomy of the proposed demonstration system. Section 4 briefly introduces the limitations of the proposed approach. In Section 5 the related work in this area is briefly discussed. In Section 6 the authors conclude and highlight some future work.

## 2 Requirements for a common interface

Ideally, manageable components and distributed applications that leverage on them through the use of business rules should be able to capitalize on the most appropriate technologies and practices for each use case. In this respect, the authors have identified four main technical challenges: a) component discovery, b) rule engine interoperability, c) rule interchangeability and d) system usability.

In order to improve the ability to manage such systems, the first step is to learn about the components and the domain it takes part of. Key knowledge includes location (e.g. Unified Resource Locator URL [8]) and types of interfaces proposed (e.g. sensors, effectors, data model discovery). As it is possible for components to be replicated, the common interface should be able to understand the relationship between instances. For instance, updating a rule for one instance should trigger the update over the others, unless specified otherwise.

With the location and list of interfaces known, it is necessary for a common interface to learn how to communicate to the rule engine. Interaction with a rule engine, for instance, necessitates the knowledge of basic rule engine operations such as how to register and unregister, execute, retrieve, filter and validate rules.

The interchangeability of rules can be divided into two parts, the exchange of the rules themselves and the acquisition of the knowledge they hold. Rule interchange requires the possibility to transform from one rule language into another. Knowledge acquisition mandates that the data be adequately described so that an external entity can comprehend it. In the current context, comprehension is based on representing data elements as metrics and their capacity, for instance, to be compared, transformed, grouped and linked.

Additionally, the common interface should expose a user-friendly view, irrespective of the languages in which facts, actions and rules provided by the underlying domains it is connected to are expressed. The usability aspect of the rule-driven approach is paramount as it allows empowering non-specialists (e.g. business individuals) with ability to design rules or manage the operation of the multi-domain system.

## 3 Anatomy of a common interface

The anatomy of a common interface introduces the demonstration system, based on the requirements described in section 2. However, in order to simplify the demonstration, in this experiment the authors assume that no semantic translation is needed. The describe work does not investigate acquisition of the knowledge between different systems.

### 3.1 Component discovery

In order to allow for the discovery and storage of the different artifacts (e.g. rule engines, translators) the authors make use of a central repository. In this experiment, the repository is implemented using the Atom Publication Protocol (APP) [9] and eXist DB [10]. Thus, different atoms feeds are used to store data about rule engines and translators. **Fig. 1.** illustrates how an atom entry is used to store data about a rule engine.

```xml
<entry xmlns="http://www.w3.org/2005/Atom">
    <id>urn:uuid:d7b3caa7-e76f-46f5-a638-16fbcc623cc9</id>
    <updated>2010-08-12T13:55:53+02:00</updated>
    <published>2010-08-12T13:55:53+02:00</published>
    <link href="?id=urn:uuid:d7b3caa7-e76f-46f5-a638-16fbcc623cc9"
        rel="edit" type="application/atom+xml" />
    <title>jess.middleware</title>
    <link href="http://some-host/jess.middleware/Functional?wsdl"
        rel="enclosure" title="functional wsdl"
        type="application/wsdl+xml" />
    <link href="http://some-host/jess.middleware/Management?wsdl"
        rel="enclosure" title="management wsdl"
        type="application/wsdl+xml" />
    <link href="http://some-host/jess.middleware/Ping?wsdl" rel="enclosure"
        title="ping wsdl" type="application/wsdl+xml" />
    <author>
        <name>Pierre de Leusse</name>
        <uri>http://home.agh.edu.pl/~pdl//</uri>
        <email>pdl@agh.edu.pl</email>
    </author>
</entry>
```

**Fig. 1.** Rule engine registration using APP

### 3.2 Rule engine interoperability

The two rule engines experimented upon present similarities that allowed the authors to design one single model of middleware interface. Thus two soap services for each rule engine are provided, allowing to control and evaluate the state of the engines' working memories. The '*Management*' service allows administration type operations and the '*Functional*' service allows operations on rules and facts in specific sets of knowledge (i.e. instance of a working memory).

**Fig. 2.** shows the different functions per web service along with the arguments they require. The WSDLs [11] used in the demonstration software do not implement all the potential functionalities and can be viewed through the web user interface.

```
Management    getProperties()
              getKnowledgeSets()
              putKnowledgeSets(List<String> knowledgeSets)
              deleteKnowledgeSets(List<String> knowledgeSets)

Functional    getRules(String knowledgeSet)
              putRules(String knowledgeSet, List<String> rules)
              deleteRules(String knowledgeSet, List<String> rules)
              validateRules(String knowledgeSet, List<String> rules)
```

**Fig. 2.** Rule engine middleware description

It is noticeable that the authors do not make the assumption that a single model of middleware is possible for every rule engines and anticipate that further experimentation will make use of different types of interfaces. The Drools project, for instance, already proposes a RESTful middleware [12].

### 3.3 Rule interchangeability

For the purpose of this experiment, the authors have chose to investigate rule interchange between Drools and Jess using the Rule Interchange Format (RIF) core language as platform neutral language. Drools and Jess were chosen for their popularity and similarities (e.g. use of Rete algorithm [13], capability to handle update of the working memory).

It is not possible to provide in this short document a full description of the translation mechanisms used in the demonstration. However, the translators themselves can be viewed and copied through the demonstration software as XSLT [14] documents.

**Fig. 3.** illustrates the different languages used to perform translations between the Drools and Jess rule engines as well as their relationships.

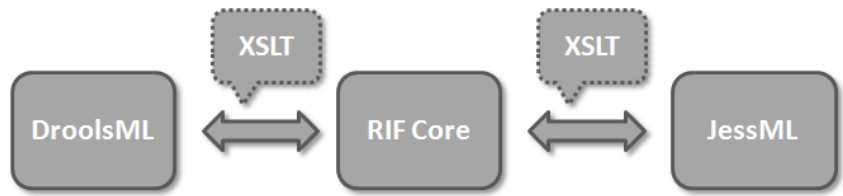

**Fig. 3.** Rule engine registration using APP

At the time of writing, the interchange soundness is verified by a specific '*validate*' function of the '*functional*' service. This function makes use of the target rule engine specific mechanism for rule validation and attempt to validate both the grammatical validity of the rule and its relevance in the current context (e.g. presence of concordant fact types in the working memory). For instance, in the case of the Jess middleware used in the demonstration system, the validations process attempts to register the rule to be validated into the working memory of the engine. Depending on the result of this operation the middleware then retrieves any error and unregisters the rule as relevant.

The authors understand the limitations of such approach and further work will investigate more appropriate techniques to evaluate the soundness of the interchange.

### 3.4  System usability

The authors have designed a web user interface for the demonstration software using Adobe Flex technology [15]. **Fig. 4**. Shows how different rule engines and their rules can be browsed together with the different functionalities available: Add rule, Delete rule, Translate rule and Modify rule (modifying a rule requires the use of the clipboard feature).

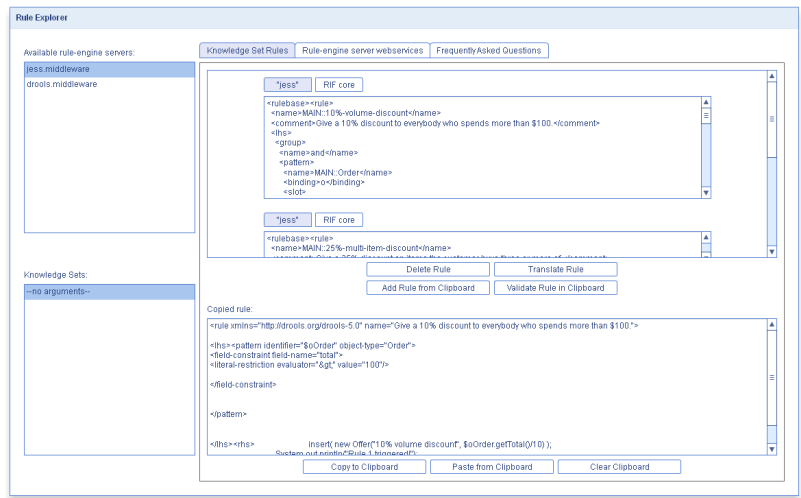

**Fig. 4.** Web user interface

The unique added value of the demonstration software's web interface at the time of writing is the ability to use different technologies of rule engines through a single user interface.

## 6  Limitations of the proposed approach

The demonstration software presents two main limitations: its interchange process and its usability.

A key requirement for this approach is the availability of reliable translations between the different rule languages and RIF core. Incremental translation can mitigate the need for specific RIF core translators. However, even assuming that the translators themselves do not have any errors, the approach suffers from the "least common denominator" limit associated with the capabilities of the supported rule engines.

The web interface, in its design at the time of writing, only proposes a textual visualization of the rules. This is not as user friendly as the editor interfaces most the commercial rule engine, Drools included, propose.

## 5 Related works

In this section, work related to the three main aspects of the experiment described in this paper and specifically related to rule-driven distributed systems are introduced. These aspects are rule engine interoperability, rule interchange and rule based system usability. The authors recognize that the domain of service discovery, although a key element, is not specific to the topic of this paper.

The main solution proposed in terms of rule engine interoperability is the JSR 94 [16] specification. JSR 94 is however specific to the Java world and does not take into account the potential distributed nature of the rule engine's environment. The other relevant piece of work related to interoperability is the service oriented business rules broker presented in [17]. The so-called business rules broker allows hiding the heterogeneity of different rules engines and providing a service-oriented interface to access and execute the business rules from different knowledge bases. The authors present the implementation of a plug-in for JSR 94 compliant rule engines. However, in [17], the authors use a XML configuration file to identify each knowledge base in one rule engine and generate a specific web service. In the current document and its demonstration software, the authors chose to separate in two services administration and functionality in order to be able to provide static interfaces instead of generated ones while enabling separation between knowledge bases.

Rule interchangeability has received a lot of interest in the past few years. Currently, the two main initiatives for rule interchangeability are RuleML [5] and RIF [6].

The RuleML Initiative develops the Rule Markup Language (RuleML). The goal of this initiative is to develop a canonical Web language for rules using XML markup, formal semantics, and efficient implementations [18]. The RuleML initiative started in 2000 and aims at becoming the standard rule markup. It already provides translations to and from different rule languages as well as other tools such as user interface.

The Rule Interchange Format (RIF) is developed by a dedicated World Wide Web Consortium (W3C) Working Group since 2005. RIF focuses on rule exchange rather than develop a specification to replace all the others. In order to address the many challenges of rules interchangeability, RIF is meant to be extensible and is divided into dialects that are all based on a common core and share as many properties as possible but each specialize in a particular domain.

As underlined in the introduction section, usability is one of the key element behind the use of business rules. It is therefore not a surprise to see that most rule engines provide rule editors. However, without providing an extensive study on user interfaces it is difficult to compare what advantages they each offer and provide a benchmark.

## 6  Conclusions and future work

In this paper and its related demonstration software, the authors have presented a common interface for multi-rule-engines distributed systems. The described work includes several novel features: firstly, a rule engine-agnostic user interface, along with its supporting tools are presented. Secondly, RIF core is used as a global language to represent rules, with translators for specific business rule engine technologies, such as Drools and Jess. Thirdly, the notion of a common interface was introduced as a means to render heterogeneous rule-driven components and the distributed systems they are part of more manageable.

Future research will focus on automated retrieval and correlation of associated rules. To allow the common interface to understand which rules or parts of rules involve the same fact and how they can be assembled to enable the end user to manipulate this fact without the need for repetitive actions.

Another domain that the authors intend to investigate is security. In particular, mechanisms are needed for rule-driven components to specify in what circumstances a user can perform certain actions on rules, facts and rule engines. For instance, allowing users to manipulate only specific parts of a rule based on the type or state of a fact should be implemented on the engine or distributed system level.

The demonstration system can be accessed at: http://home.agh.edu.pl/~bkwolek/yield/ alternatively, the reader can see a video of the demonstrated system at: http://home.agh.edu.pl/~bkwolek/yield/RuleML_demo.avi.

## Acknowledgements

The authors acknowledge Joanna Kosinska, Robert Szymacha, Marek Psiuk, Daniel Żmuda, Tomasz Szydlo and Sławek Zieliński from the Distributed System Research Group, AGH University of Science and Technology for their significant contribution to the development and improvement of the work presented in this paper.

This work is part of the IT SOA project founded by the European Union and the Polish Minister Of Higher Education. More details on this project can be found at: http://www.soa.edu.pl